\documentclass[12pt,amsfonts]{article}    

\usepackage{amsfonts}

\def\one{1\hskip-.37em 1}
\def\Oh{{\cal{O}}} 

\def\vp{\varphi}
\def\bl{\mathbf{\l}}

\def\half{\textstyle{\frac{1}{2}}}

\def\H{{\cal H}}

\def\L{\Lambda}

\def\p{\varphi}

\def\th{\theta}
\def\H{{\cal H}}
\def\B{\beta}

\def\l{\lambda}

\def\si{\sigma}

\def\t{\textstyle}

\def\ra{\rightarrow}
\def\tint{{\textstyle\int}}

\def\d{\partial}

\def\b{\begin{eqnarray*}}  
\def\e{\end{eqnarray*}}    
\def\bn{\begin{eqnarray}}  
\def\en{\end{eqnarray}}   

\def\<{\langle}
\def\>{\rangle}

\def\bk{\mathbf k}
\def\bm{\mathbf m}

\def\de{\delta}
\def\no{\nonumber}

\def\ds{d^s\!x}
\def\k{\kappa}
\def\bl{\mathbf l}

\def\hk{\hat{\kappa}}
\def\{{\lbrace}
\def\hv{\hat{\varphi}}
\def\}{\rbrace}
\begin{document}

\title{Using Coherent States to Make Physically \\Correct Classical-to-Quantum Procedures that\\
Help Resolve Nonrenomalizable  Fields \\Including Einstein's Gravity}
  \author{John R. Klauder\footnote{klauder@ufl.edu} \\
Department of Physics and Department of Mathematics  \\ 
University of Florida,   
Gainesville, FL 32611-8440}
\date{ }
\bibliographystyle{unsrt}

\maketitle 

\begin{abstract}
Canonical quantization covers a broad class of classical systems, but that does not include all the problems of interest. Affine quantization has the benefit of providing a successful quantization of many important    
problems including the quantization of half-harmonic oscillators \cite{11},
nonrenormalizable scalar fields, such as $(\vp^{12})_3$ \cite{1} and $(\vp^4)_4$ \cite{2}, as well as  the quantum theory of Einstein's general relativity \cite{22}. The features that distinguish affine quantization are emphasized, especially, that affine quantization differs from canonical quantization only by the choice of classical variables promoted to quantum operators. Coherent states are used to ensure proper quantizations are physically correct.  While quantization of nonrenormalizable covariant scalars and gravity are difficult, we focus on appropriate ultralocal scalars and gravity which are
fully soluble while, in that case, implying that affine quantization is the proper procedure to ensure
the validity of affine quantizations for nonrenormalizable covariant scalar fields and Einstein's gravity.          
\end{abstract}

\section{Introduction}
We first begin with three basic quantization procedures which include Sec.~1.1 canonical quantization, Sec.~1.2 spin quantization,
and Sec.~1.3 affine quantization. For a particular system, there are many correct classical variables, but, basically, only one set of these classical variables leads to valid quantum variables. For each procedure we are led to certain rules that guarantee a physically correct quantization. The rules we present should precede any model's analysis in order to ensure a properly physical result.

\subsection{A brief review of canonical quantization}
Classical variables $p\&q$ that obey $-\infty<p,q<\infty$ and have a Poisson bracket $\{q.p\}=1$
are candidates to promote to basic quantum operators $P\& Q$, which obey $[Q, P]=i\hbar 1\!\!1$.
For convenience we choose $q\&Q$ as dimensionless, while then $p\& P\& \omega$ ($\omega$ appears below) have the dimensions of $\hbar$.
However, $P\& Q$ will be physically correct operators {\it provided} that the original variables $p\& q$ were `Cartesian coordinates' \cite{6}. 

\subsubsection{Canonical coherent states}
Cartesian coordinates can be found in normalized coherent states of the form
$|p,q\>\equiv e^{-iqP/\hbar} e^{ipQ/\hbar}\,|\omega\>$ with $(Q+iP/\omega)|\omega\>=0$, which implies that $\<\omega|Q|\omega\>=\<\omega|P|\omega\>=0$. For any operator expression, like $\H(P,Q)$, the coherent states lead to 
\bn \<p,q|\H(P,Q)|p,q\> =\<\omega| \H(P+p, Q+q)|\omega\>=H(p,q) +\Oh(\hbar;p, q) \;. \en
The $H(p,q)$ term is free of $\hbar$, which implies that the $\H(p,q)=H(p,q)$ as Dirac also required \cite{6}. Moreover, the whole line of (1) is {\it independent of any phase factor} of the coherent states, such as $|p, q: f\>= e^{i f(p, q)}|p,q\>$. This independence is carried over to a Fubini-Study metric \cite {7}, namely
  \bn d\sigma(p,q)^2\equiv 2\hbar\,[ |\!|\,d|p,q\>|\!|^2 -|\<p,q|\,d|p,q\>|^2] =\omega^{-1}\,dp^2 + \omega\,dq^2 \;, \en
  which leads us to suitable Cartesian coordinates! More generally, this two-dimensional space may
   be called a  `constant zero curvature' surface. It is noteworthy that this `constant zero curvature' was not {\it sought}, it was {\it created!} Efforts to use canonical quantization with
   classical variables that do not belong to a `constant zero curvature' are very likely to lead to a non-physically correct quantization.  
  
\subsection{A brief review of spin quantization}
The operators in this story are $S_i$ with $i=1,2,3$, and which (here $i=\sqrt{-1}\,$) satisfy $[S_i, S_j]=i\,\hbar\,\epsilon_{ijk} \, S_k$ These operators obey $\Sigma_{l=1}^3\,S_l^2=\hbar^2 s(s+1)1\!\!1_{2s+1}$, where $2s+1=2,3,4,...$ is the dimension of the spin matrices. The normalized eigenvectors of $S_3$ are $ S_3|s,m\> =m\hbar|s,m\>$, where $m\in\{-s,...,s-1,s\}$.

\subsubsection{Spin coherent states}
The spin coherent states are defined by
   \bn |\th,\vp\>\equiv e^{-i\vp S_3/\hbar}\,e^{-i\th S_2/\hbar}\,|s,s\>\;, \en
where $-\pi<\vp\leq\pi$, and $0\leq\th\leq\pi$. 
It follows that
  \bn &&d\si(\th,\vp)^2 \label{ff} \\
  &&\hskip1em \equiv 2\hbar\,[\,|
  \!|\,d|\th,\vp\>|\!|^2-|\<\th,\vp|\,d|\th,\vp\>|^2\,] 
  =(s\hbar)[d\th^2+\sin(\th)^2\,d\vp^2\,
  ]\;,\no \en
  We can also introduce $q=(s\hbar)^{1/2}\,\vp$
and $p=(s\hbar)^{1/2}\,\cos(\th)$, along with $|p,q\>=|p(\th,\vp),q(\th,\vp)\>$,
  which leads to
  \bn &&\hskip-2em d\si(p,q)^2 \label{ss} \\
  &&\hskip-1em \equiv 2\hbar\,[\,|\!|\,d|p,q\>|\!|^2-|\<p,q|\,d|p,q\>|^2\,] 
  =(1-p^2/s\hbar)^{-1}
  dp^2+(1-p^2/s\hbar)\,dq^2\;. \no \en
  Equation (\ref{ff}) makes it clear that we are dealing with a spherical surface with a radius of $(s\hbar)^{1/2}$; this space is also known as a `constant positive curvature' surface, and it has been created! These classical variables can not lead to a physically correct canonical quantization. Instead, they offer a distinct quantization procedure that applies to different problems. However,
  Eq.~ (\ref{ss}) makes it clear that if $s\ra\infty$, in which case both  $p$ and $q$ span the real line, we are led to `Cartesian coordinates', a basic property of canonical quantization.
  
\subsection{A brief review of affine quantization}
Consider a classical system for which $-\infty< p <\infty$, but $0<q<\infty$, that does not lead to self-adjoint quantum operators. Perhaps we can do better if we change classical variables. For example, 
 the classical action factor $p\,dq= pq \,dq/q=pq\,d\ln(q)$, leads to proper variables to promote to quantum operators. In particular, $pq\ra(P^\dag Q+QP)/2 \equiv D \;(=D^\dag)$. However, besides 
 $0<q<\infty$, it may arise that $-\infty<q<0$, or even $-\infty<q\neq0<\infty$ (e.q., if $q^{-2}$ is part of the potential).
 To capture all three possibilities for $q$ -- and thus also for $Q \,( = Q^\dag)$ -- we are led to  $[Q,D]=i\hbar Q$, which happens to be the Lie algebra of the ``affine group'' \cite{3}, and,
 incidentally, gives its name to affine quantization. 
 Again, it is useful to choose dimensions such that $q \& Q$ are dimensionless while $p \& D$ have the dimensions of $\hbar$.
 
 \subsubsection{Affine coherent states}
 The affine coherent states involve the quantum operators $D$ and now $Q>0$, and we use the classical variables $p$ and $\ln(q)$, with $q>0$. Specifically, we choose
     \bn |p;q\>\equiv e^{ipQ/\hbar} \,e^{-i \ln(q)\,D/\hbar}\,|\B\>\;, \en
  where the fiducial vector $|\B\>$ fulfills the condition $ [(Q-1\!\!1)+iD/\B]|\B\>=0$, which implies that
  $\<\B|Q|\B\>=1$ and $\<\B|D|\B\>=0$.\footnote{The semicolon in $|p;q\>$ distinguishes the affine ket from the canonical ket  $|p,q\>$. If $-\infty<q<0$,
  or $-\infty<q\neq0<\infty$, then replace  the fiducial vector, change  $\ln(q)$ to $\ln(|q|)$, and keep $q\ra Q$.}This expression leads to
  \bn \<p;q|\H'(D,Q)|p;q\>=\<\B|\H'(D+pqQ,qQ)|\B\> = H'(pq,q) +\Oh'(\hbar; p, q) \;,\en
  and, as $\hbar\ra0$, $\H'(pq,q)=H'(pq,q)$ as Dirac \cite{6} has required.
  It follows that the Fubini-Study metric, for $q>0$, becomes
   \bn  &&d\sigma(p;q)^2 \\ 
   &&\hskip1em \equiv 2\hbar[|\!|\, d|p;q\>|\!|^2 -|\<p;q|\,d|p;q\>|^2]=\B^{-1}q^2\,dp^2+ \B\,q^{-2}\,dq^2 \;. \no \en
   This expression leads to a surface that has a `constant negative curvature' \cite{ss} of magnitude $-2/\B$, which, like the other curvatures, 
   has been `created'. This set of classical variables can not lead  to a physically correct
   canonical quantization.    Instead, they offer a distinct quantization procedure that applies to different problems. Any use of classical variables that do not form a `constant negative curvature' subject to an affine quantization is very likely not a physically correct quantization.
   
The rule that $0<q<\infty$ is limited and we can easily consider $0< q+k <\infty$, where $k>0$. This changes the coherent states from $\ln(q)$ to $\ln(q+k)$, which then changes the Fubini-Study metric to 
 $\B^{-1} (q+k)^2\, dp^2+ \B\,(q+k)^{-2} \, dq^2$. If we choose to let $k\ra \infty$ and at the same time let $\B\ra (\B+\omega k^2)$, we are led to $\omega ^{-1}dp^2+\omega \,dq^2$, now with $q\in I\!\!R$, which, once again, applies to canonical quantization.
 
  These three stories complete our family of `constant curvature' spaces. Additionally, the various coherent states can build ``bridges'' in each case from the classical realm to the quantum realm \cite{j2,j3}.

 \section{Affine Quantization of \\Nonrenormalizable Scalar Fields}
Our goal here is to consider the quantization of scalar fields such as the classical action functional given by
  \bn A_c=\tint_0^T\tint \{(1/2)[\dot{\vp}(t,x))^2 -(\overrightarrow{\nabla}\vp)(t,x)^2 -m_0^2\,\vp(t,x)^2] - 
  g\,\vp(t,x)^r\}\;d^s\!x\,dt \en
  for selected examples of the power $r>2$ and where $s$ denotes the number of spatial dimensions. We also define $n= s+1$ as the total number of spatial and temporal dimensions.
  
  Canonical quantization is successful for $r<2n/(n-2)$, but is not successful for $r\geq2n/(n-2)$.
The reason for the failures in the latter case can be traced to the change of a classical domain that is required  in the latter case that forces the usual free solution to be {\it disconnected} from any non-free solution. We will illustrate that failure in a much ``simpler model'' to make it clear. This simpler model is fully soluble and does {\it not} include a typical free solution as one if its solutions! Moreover, the solutions of the simple model will simultaneous ensure that {\it affine quantization can solve nonrenormalizable covariant scalar fields as well}.\footnote{The following sections are partially based on \cite{98}.}

       \subsection{A regularized affine ultralocal scalar field}
       Our regularization is of the underlying space in which $x\ra \bk\,a$, where $\bk\in\{..., -1,0,1,2,3,...\}^s$ and $a>0$ denotes the tiny distance between lattice rungs. The regularized 
       classical ultralocal (= NO gradients) Hamiltonian is given, for $r>2$ and $s\geq1$, by
         \bn H_u={\t\sum}_{\bk} \{\half[\pi_\bk^2+ m_0^2\,\vp_\bk^2]+g_0\,\vp_\bk^r\}\;a^s \;.\en
         The classical affine regularization involves $\k_\bk=\pi_\bk\,\vp_\bk$ and $\vp_\bk$, with
       $\vp_\bk\neq0$, and a Poisson bracket $\{\vp_\bk, \k_\bm\}=\delta_{\bk,\bm}\,\vp_\bk$. This  leads to the classical affine regularized ultralocal Hamiltonian given by
   \bn H'_u= {\t\sum}_{\bk} \{\half[\k_\bk^2\vp_\bk^{-2}+m_0^2\,\vp_\bk^2]+g_0\,\vp_\bk^r\}\;a^s \;.\en
   
 The regularized basic quantum Schr\"odinger operators are given by $\hv_\bk=\vp_\bk\neq0$ and
           \bn &&\hk_\bk=-i\half\hbar[ \vp_\bk(\d/\d\vp_\bk)+(\d/\d\vp_\bk)\vp_\bk] a^{-s} \no \\
           &&\hskip1.36em=-i\hbar[\vp_\bk(\d/\d\vp_\bk)+1/2]a^{-s} \;. \en
           An important result is that $\hk_\bk\,\vp_\bk^{-1/2}=0$. The role of the Schr\"odinger equation becomes
  \bn i\hbar \,\d\psi(\vp,t)/\d t={\t\sum}_\bk\{\half[\hk_\bk\vp_\bk^{-2}\hk_\bk+m_0^2\,\vp_\bk^2]+g_0\,\vp_\bk^r \}\;a^s\;\psi(\vp,t)\;. \en
  The normalized ground state of such an equation may, with $b\approx 1$ and $ba^s$ dimensionless, be given by
      \bn \psi_0(\vp)= \Pi_\bk e^{-V(\vp_\bk)/2}\,(ba^s)^{1/2}\,\vp_\bk^{-(1-2ba^s)/2}\;, \en
      for some real function $V(\vp_\bk)$.\footnote{The expression $\hk_\bl \,\Pi_\bk \,\vp_\bk^{-1/2}=0$, with $\vp_\bk\neq0$, is an analog to $\pi(x)\,1\!\!1=0$, which is self evident. This
      accounts for the term $\Pi_\bk \vp_\bk^{-1/2}$ as part of the vector states.}
      Finally, we ask what is the characteristic function for such an equation, and the answer is given by
       \bn &&C(f)=\lim_{a\ra0}\Pi_\bk\,\tint \;e^{i f_\bk \vp_\bk}\; e^{-V(\vp_\bk)}\,(ba^s)\,|\vp_\bk|^{-(1-2ba^s)}\;d\vp_\bk \no \\
       &&\hskip2.53em  =\lim_{a\ra0}\Pi_\bk \{1-(ba^s)\tint[1-e^{if_\bk\vp_\bk }]\,e^{-V(\vp_\bk)}\,|\vp_\bk|^{
       -(1-2ba^s)}\;d\vp_\bk \} \no \\
       &&\hskip2.55em  =\exp\{-b\tint \ds\tint[1-e^{i\,f(x)\l}] e^{-v(\l)}\,d\l/|\l|\} \;.\label{x}\en
       Here $\vp_\bk\ra\l$, and $V\ra v$ to account for changes that may have arisen in $V$ as $a\ra0$. {\bf Note:} For other solutions replace $e^{-v(\l)}$ by $|w(\l)|^2$ for  a suitable function $w(\l)$.
       
       When $g_0\ra0$ this characteristic function does {\it not} become a Gaussian because
       the domain of allowed functions has become smaller than had been present when one starts with
       $g_0\equiv 0$.
       The resultant expression in (\ref{x}) is a (generalized) Poisson distribution, which, besides 
       a Gaussian distribution, is the only other form allowed by the Central Limit Theorem \cite{lt}.

       \subsubsection{The main lesson from ultralocal scalar fields}
       The previous subsection found that an ultralocal scalar field model led to acceptable results 
       when $r>2$ and $n\geq2$. For certain covariant scalar field models, we have already 
       observed that acceptable results arise by canonical quantization when  $r<2n/(n-2)$. 
       In view of acceptable results for ultralocal scalar fields when $r>$ and $n\geq2$,
       we predict that an affine quantization for covariant scalar fields leads to acceptable results 
       when $r\geq 2n/(n-2)$. 
       
       Monte Carlo studies, such as those carried out in $(\vp^{12})_3$ \cite{1} and $(\vp^4)_4$
       \cite{2}, have confirmed that affine quantization of nonrenormalizable models can lead to physically acceptable results. 
       
       \section{Affine Quantization of General Relativity}
        An effort to quantize Einstein's theory of gravity has been examined in several
        articles published by the author; see \cite{22,4,5,8}. In light of those articles, we will present a modest selection 
        of the necessary features for an affine quantization of Einstein's gravity.
        
        The ADM classical Hamiltonian \cite{adm} with $g(x)\equiv \det[g_{ab}(x)]>0$, is given by
          \bn H_c=\tint\{ g(x)^{-1/2}[\pi^a_b(x)\pi^b_a(x)-\half\,\pi^a_a(x)\pi^b_b(x)]+
          g(x)^{1/2}\,R(x)\,\}\;d^3\!x\; \en
       where $\pi^a_b(x)\equiv \pi^{ac}(x)g_{bc}(x)$, and $R(x)$ is the 3-dimensional scalar curvature. 
       
       The term $R(x)$ contains the spatial derivatives and the ultralocal version of the 
       classical Hamiltonian is chosen as
        \bn H_u=\tint\{ g(x)^{-1/2}[\pi^a_b(x)\pi^b_a(x)-\half\,\pi^a_a(x)\pi^b_b(x)]+
          g(x)^{1/2}\,\Lambda(x)\}\;d^3\!x\;, \en
          where $\Lambda(x)$ is a fixed, spatially dependent, continuous function that takes the place of the  scalar  curvature. When quantized, the only variables that are promoted
          to quantum operators are the metric field, $g_{ab}(x)$, and the momentric 
          field, $\pi^c_d(x)$.

          \subsection{A regularized affine ultralocal quantum gravity}
          Much like the regularization of the scalar fields, we introduce a discrete version of the 
          underlying space such as $x\ra \bk a$, where $\bk\in\{...,  -1,0,1,2,3,...\}^3$ and
          $a>0$ is the spacing between rungs in
          which, for the Schr\"odinger representation, $g_{ab}(x)\ra g_{ab\,\bk}$ and 
          $\hat{\pi}^c_d(x)\ra \hat{\pi}^c_{d\,\bk}$ that becomes
             \bn &&\hat{\pi}^c_{d\,\bk}=-i\half\hbar[ g_{de\,\bk} (\d/\d g_{ce\,\bk})+
              (\d/\d g_{ce\,\bk})g_{de\,\bk}]\;a^{-3}\;\no \\
              &&\hskip1.9em
               =-i\hbar[g_{de\,\bk}(\d/\d g_{ce\,\bk})+\delta^c_d/2]\;a^{-3}\:.\en
               Take note that $\hat{\pi}^a_{b\,\bk}\,g_\bk^{-1/2}=0$, 
               where $g_\bk\equiv \det(g_{ab\,\bk})>0$.
               
               The regularized Schr\"odinger equation is then given by
               \bn i\hbar\,\d \psi(\{g\},t)/\d t={\t\sum}_\bk\{\hat{\pi}^a_{b\,\bk}g_\bk^{-1/2}
               \hat{\pi}^b_{a\,\bk}-\half \hat{\pi}^a_{a\,\bk}g_{\bk}^{-1/2}\hat{\pi}^b_{b\,\bk}\no\\
               +g_\bk^{1/2}\Lambda_\bk\,\}\;a^3\;\psi(\{g\},t)\;,\en
               where $\{g\}\equiv \{g_{ab \,\bk}\}$.
               A normalized, stationary solution to this equation is given by
               \bn \psi_Y(\{g\})=\Pi_\bk\,Y(g_\bk,\L_\bk)\,(ba^3)^{1/2}\,
               g_\bk^{-(1-ba^3)/2}\;.\en
               The characteristic function for such an expression is given by
             \bn &&C_Y(f)=\lim_{a\ra0}\Pi_\bk\tint e^{if_\bk g_\bk}\,|Y(g_\bk,\L_\bk)|^2 (ba^3)
             g_\bk^{-(1-ba^3)}\:dg_\bk\no\\
              &&\hskip3em=\lim_{a\ra0}\Pi_\bk\{1-(ba^3)\tint[1-e^{if_\bk g_\bk}]|Y(g_\bk,\L_\bk)|^2
              g_\bk^{-(1-ba^3)}\;dg_\bk \} \no \\
      &&\hskip3em=\exp\{-b\tint d^3\!x\,\tint[1-e^{if(x)\mu}]\,|y(\mu, \L(x))|^2\,d\mu/\mu \}\;, \en
             where the scalar $g_\bk\ra\mu>0$ and $Y\ra y$ to accommodate
             any change in $Y$ due to $a\ra0$. Once again, the final result is a (generalized) Poisson 
             distribution, which obeys the Central Limit Theorem \cite{lt}.

               \subsubsection{The main lesson from ultralocal gravity}
               Just like the success of quantizing ultralocal scalar models, we have also showed that 
               ultralocal gravity  can be quantized using affine quantization. The purpose of solving 
               ultralocal scalar models was to
               ensure that non-renormalizable covariant scalar fields can also be solved using affine 
               quantization. Likewise, the purpose of quantizing an ultralocal version of Einstein's
               gravity shows that we should, in principle, and using affine quantization, be able 
               to quantize the genuine version of Einstein's gravity using affine quantization.  Specific details pointing toward Einstein's gravity are presented in \cite{22}.
               
               Monte Carlo evaluations have begun for nonrenormalizable covariant scalar fields
               in \cite{1,2} using affine quantization with acceptable results. 
               Perhaps, some Monte Carlo evaluations may also be used to examine some quantum
               features of Einstein's gravity using affine quantization.

\end{document}